\documentclass[twocolumn]{aastex7}
\usepackage{threeparttable}
\usepackage{booktabs}
\usepackage{url}

\newcommand{\rev}[1]{\textcolor{black}{#1}}

\begin{document}

\title{Intense Hard X-ray Emissions in C-class Flares: A Statistical Study with ASO-S/HXI Data}

\author{Changxue Chen}
\affil{Key Laboratory of Dark Matter and Space Astronomy, Purple Mountain Observatory, Chinese Academy of Sciences, Nanjing 210023, People’s Republic of China}
\affil{School of Astronomy and Space Science, University of Science and Technology of China, Hefei 230026, People’s Republic of China}
\email{cxchen@pmo.ac.cn}

\correspondingauthor{Yang Su}
\email{yang.su@pmo.ac.cn}
\author[0000-0002-4241-9921]{Yang Su}
\affil{Key Laboratory of Dark Matter and Space Astronomy, Purple Mountain Observatory, Chinese Academy of Sciences, Nanjing 210023, People’s Republic of China}
\affil{School of Astronomy and Space Science, University of Science and Technology of China, Hefei 230026, People’s Republic of China}
\email{yang.su@pmo.ac.cn}

\author{Wei Chen}
\affil{Key Laboratory of Dark Matter and Space Astronomy, Purple Mountain Observatory, Chinese Academy of Sciences, Nanjing 210023, People’s Republic of China}
\email{w.chen@pmo.ac.cn}

\author{Jingwei Li}
\affil{Key Laboratory of Dark Matter and Space Astronomy, Purple Mountain Observatory, Chinese Academy of Sciences, Nanjing 210023, People’s Republic of China}
\email{nj.lijw@pmo.ac.cn}

\author{Fu Yu}
\affil{Key Laboratory of Dark Matter and Space Astronomy, Purple Mountain Observatory, Chinese Academy of Sciences, Nanjing 210023, People’s Republic of China}
\email{fuyu@pmo.ac.cn}

\author{Weiqun Gan}
\affil{Key Laboratory of Dark Matter and Space Astronomy, Purple Mountain Observatory, Chinese Academy of Sciences, Nanjing 210023, People’s Republic of China}
\email{wqgan@pmo.ac.cn}

\begin{abstract}
In the standard model of solar eruptive events, coronal mass ejections (CMEs) and flares are associated with each other through magnetic reconnection initiated by erupting flux ropes. Observations also reveal an increasing association ratio between flares and CMEs with flare intensity. However, the fundamental relationship between flares and CMEs, and that between thermal and nonthermal processes, remains unknown.
Here we investigate energetic C-class flares (ECFs)---Geostationary Operational
Environmental Satellite (GOES) C-class flares with hard X-ray (HXR) emissions above 30 keV---using observations from Advanced Space-based Solar Observatory/Hard X-ray Imager (HXI), \rev{Solar Dynamic Observatory}, and GOES. Among 1331 C-class flares detected by HXI, 127 ECFs (9.5\%) were identified for statistical analysis of their properties and associations with CMEs and other flare-related features. 
Our statistical results reveal that ECFs have relatively shorter durations and harder spectra (the mean electron \rev{power-law} index is 4.65), with no significant correlation between soft X-ray flux and nonthermal parameters (e.g., HXR peak flux).
Among the 127 events, 53 (42\%) were associated with type III bursts, 38 (30\%) with jets, at least 13 ($\sim$11\%) with \rev{360 nm brightenings}, and only 5 ($\sim$4\%) with CMEs.
Crucially, all five CME events were narrow CMEs associated with jets. The surprising correlation between these ECFs and CMEs suggests that noneruptive or confined magnetic field configurations in these flares may favor electron acceleration, resulting in harder X-ray spectra.
We \rev{discuss} the potential formation mechanisms and efficient electron acceleration processes in these atypical flares, providing valuable insights into nonstandard flare behavior.
 
\end{abstract}


\section{Introduction} \label{sec:intro}

Accelerated particles in solar flares carry a significant amount of flare energy \citep{1995ARA&A..33..239H}, which is important for understanding energy release and plasma heating in flares. The acceleration and propagation processes reflect the underlying physics of flares. A portion of the accelerated electrons may travel outward along ``open'' magnetic field lines into interplanetary space, generating type III radio bursts through the resonant excitation of Langmuir waves \citep{radio2014RAA....14..773R}.
The remaining electrons propagate downward along magnetic field lines, depositing energy into the solar chromosphere. This energy deposition produces hard X-ray (HXR) emissions via nonthermal bremsstrahlung \citep{1971SoPh...18..489B,holmanImplicationsXrayObservations2011} and causes intense heating, which makes the heated plasma observable in the soft X-ray (SXR) band. 
The enormous thermal pressure generated by this heating drives the high-temperature plasma upward along magnetic field lines to fill the flare loops, a process known as ``chromospheric evaporation'' \citep[see the review of][]{fletcherObservationalOverviewSolar2011b}. 
The Neupert effect \citep{neupertComparisonSolarXRay1968,veronigInvestigationNeupertEffect2002}, in which the cumulative HXR flux from the accelerated electrons matches the SXR flux produced by the heated plasma, provides indirect evidence for chromospheric evaporation.
In particularly strong flares, white-light (WL) emission produced either by direct electron-beam heating of a deeper atmosphere \citep[e.g.,][]{1972SoPh...24..414H,kruckerCOSPATIALWHITELIGHT2015} or by secondary effects, such as the ``radiative backwarming'' model \citep[e.g.,][]{Backwarming1989SoPh..124..303M,dingInterpretationInfraredContinuum2003} may be observed.

Based on the theory of chromospheric evaporation, both the thermal and nonthermal radiations in flares are associated with accelerated electrons, and thus they are typically correlated in statistical studies. For example, \citet{battagliaSizeDependenceSolar2005} and \citet{2008SoPh..250...53S} found positive correlations between the 1--8 $\mathrm{\AA}$ SXR flux and the 35 and 50 keV HXR fluxes, respectively. In addition, \citet{warmuthConstraintsEnergyRelease2016a,warmuthConstraintsEnergyRelease2016} analyzed 24 flares ranging from C- to X-class and discovered positive correlations among the nonthermal electron energy flux, SXR flux, and plasma thermal energy.

However, this correlation is not absolute. Some flares are entirely dominated by thermal plasma, with almost no nonthermal radiation detected \citep{battagliaObservationsConductionDriven2009,fleishmanENERGYPARTITIONSEVOLUTION2015}, while in \rev{some} other flares, strong nonthermal \rev{HXR} radiation is observed with \rev{little or nearly no thermal response, as seen in early impulsive flares \citep{suiNonthermalXRaySpectral2007}, ``cold" flares \citep{fleishmanCOLDTENUOUSSOLAR2011,lysenkoStatisticsColdEarly2018} and hard microflares \citep{2024A&A...683A..41S,2024A&A...691A.172B}}. Consequently, the thermal and nonthermal components in flares can exist in arbitrary proportions, and systematic studies of flares under various conditions are essential to deepen our understanding of flare energy transport processes.

Studies showing a positive correlation between HXR and SXR fluxes have mainly relied on samples of eruptive flares \citep[e.g.,][]{2008SoPh..250...53S,warmuthConstraintsEnergyRelease2016a}. These events follow the standard flare model \citep{aprocessforflare1964,sturrockModelHighEnergyPhase1966,hirayamaTheoreticalModelFlares1974,koppMagneticReconnectionCorona1976}, where magnetic reconnection is initiated by the ejection of magnetic flux ropes (MFRs), typically resulting in an accompanying coronal mass ejection (CME). In contrast, hard microflares are generally manifested as only a few brightening loops and not associated with the CME, clearly diverging from the standard model. Nevertheless, they still exhibit hard spectra, indicating strong nonthermal radiation from high-energy electrons, a characteristic more commonly seen in large eruptive flares.
On the other hand, previous statistical studies of the flare–-CME relationship have found that the correlation between the two increases with flare magnitude. For example, \citet{yashiroVisibilityCoronalMass2005} reported that only about 20\% of C-class flares are accompanied by CMEs, while all flares above X3 are associated with a CME, suggesting that lower-class flares increasingly deviate from the eruptive process described by the standard model. This raises the question of whether the efficient electron acceleration process in these lower-class flares is related to their departure from the standard model.

Given that hard microflares, characterized by classes below C2 and nonthermal electron \rev{power-law} indices under 5, represent an extreme case in the thermal-to-nonthermal ratio, are very rare (with an occurrence rate of about 1.5\% \citep{hannahRHESSIMicroflareStatistics2008}), and are predominantly compact and confined events, they are not ideal for testing this hypothesis. Therefore, in this Letter, we focus on a statistical analysis of energetic C-class flares (ECFs) that exhibit strong HXR emissions observed by the Hard X-ray Imager \citep[HXI;][]{zhangHardXrayImager2019,suTestsCalibrationsHard2024} on board the Advanced Space-based Solar Observatory \citep[ASO-S;][]{ganChineseSolarObservatory2022a,ganAdvancedSpaceBasedSolar2023}, aiming to understand the similarities and differences between this type of flare and ``normal'' C-class flares (NCFs), as well as whether they involve distinct energy transfer processes. The sources of statistical data and the process of data filtering and processing are described in Section~\ref{sec:method}.
In Section~\ref{sec:result}, we present the statistical distribution and correlations of various parameters of ECFs. Finally, in Section~\ref{sec:summary}, we summarize and discuss key observational characteristics.

\section{Data Reduction and Processing} \label{sec:method}
The solar flare magnitude (A- to X-class) is typically defined according to the SXR 1--8 $\mathrm{\AA}$ irradiances observed by the X-ray Sensor \citep{1996SPIE.2812..344H} on board the Geostationary Operational Environmental Satellite (GOES).
This study utilizes solar HXR observations provided by the HXI, a spatially modulated Fourier transform X-ray imager that captures X-ray light curves, spectra, and images in the energy range of 10--300 keV from Earth’s perspective.
The finest spatial resolution is as high as $3.1^{\prime\prime}$ and the time resolution is 4 s in normal mode and can be as high as 0.125 s in burst mode.
Note that the satellite is in a Sun-synchronous orbit \citep[720 km;][]{ganAdvancedSpaceBasedSolar2023}, so \rev{it periodically flies} through the South Atlantic Anomaly (SAA) and Earth's radiation belt. The flux enhancement generated \rev{by the particles in these regions} does not correspond to the solar flares and needs to be carefully distinguished.

\subsection{Flare Detection}

\begin{figure}[!htbp]
	\centering
	\includegraphics[width=1\textwidth]{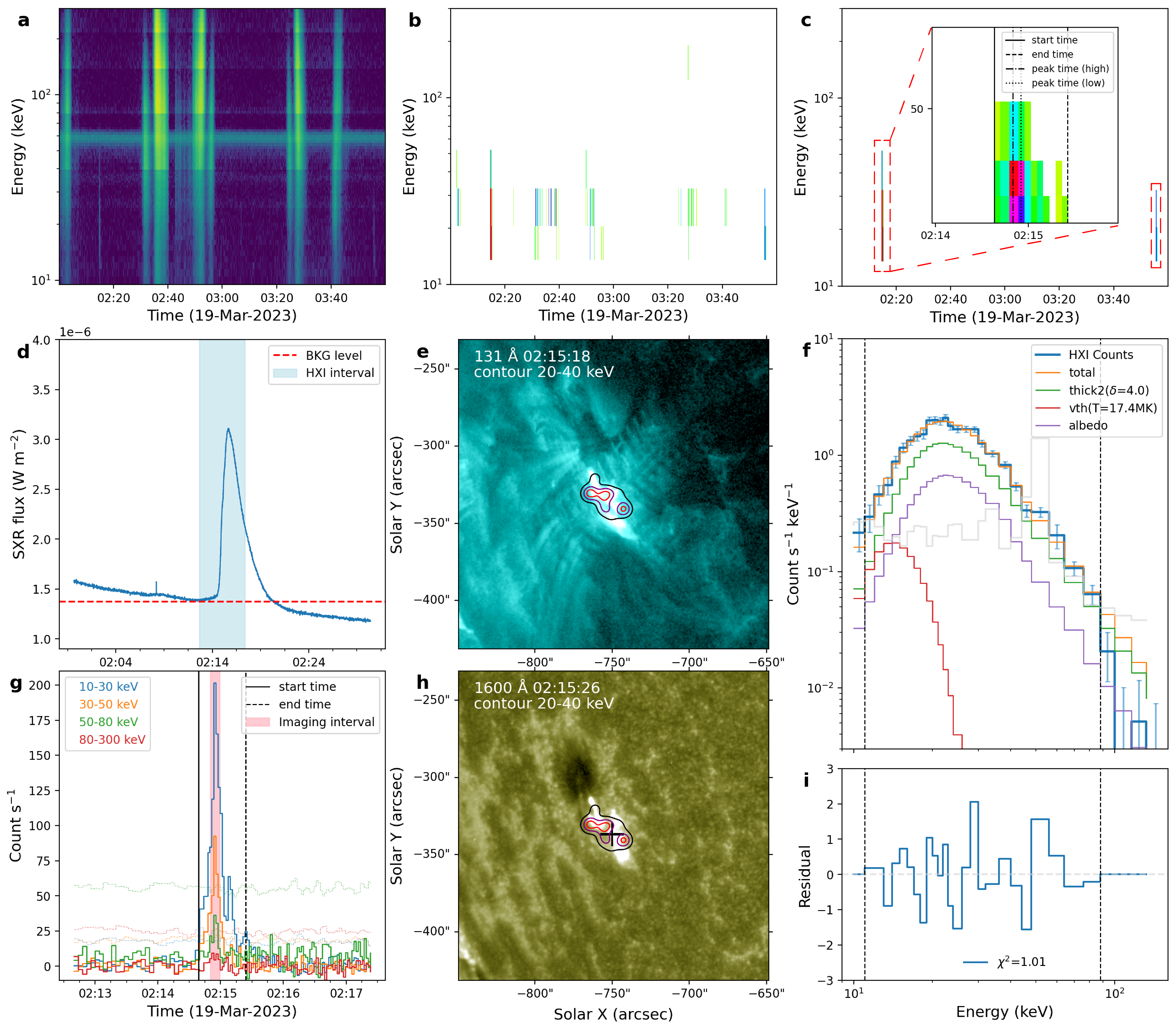} 

    \caption{HXI flare identification process and data processing. (a) HXI raw spectrogram in 80 energy bands. It should be noted that many of the enhanced emissions are not originating from flares, but from particles in Earth's radiation belts. (b) The background-subtracted spectrogram. For computational simplification, the data were binned into nine energy bins, with only pixels above the threshold presented. (c) Flare spectrogram, with pixels in the image derived from those in panel (b) that satisfy both connectivity and flux thresholds, corresponding to genuine flare-induced emission enhancements. During the displayed time period, two flares were identified (annotated with red dashed boxes), with the details of one shown in the magnified inset. (d)--(i) The processing flow of SOL2023-03-19T02:15. (d) SXR light curve, with the blue shaded area representing the time range of the panel (g) image and the red dashed line representing the SXR background level we have selected. (e) Observation of the flare in the AIA 131 $\mathrm{\AA}$, with superimposed contour lines in red (80\% of peak intensity), purple (50\%), and black (20\%) representing HXR 20–-40 keV emission. (f), (i) The spectrum fitting results (including residuals) near the peak time for 10 s, where different fitting components are represented by different colored lines and displayed in the legend, and the gray lines represent the background level. (g) HXR light curves, with dashed lines of the same color representing the background level of each energy channel. The vertical solid and dashed lines represent the start and end times of flare identification, while the red shaded areas represent the time periods corresponding to the imaging in panels (e) and (h). (h) Similar to panel (e), but the background image shows observations in AIA 1600 $\mathrm{\AA}$, where the centroid corresponding to the flare ribbons is marked with a black plus sign.} 
 
    \label{fig:img1}
\end{figure}

We initiate our analysis by detecting solar flares from HXI data. The flare identification procedure entails three primary stages (as shown in Figure~\ref{fig:img1}(a)--(c)):
\begin{enumerate}
    \item We reduce the Level 1 data (Figure~\ref{fig:img1}(a)) from the total flux detector D94 and its adjacent background detector D99 by merging temporal bins into 4 s intervals and 80 energy channels into nine consolidated bands: 10--12, 12--20, 20--30, 30--50, 50--80, 80--120, 120--180, 180--284, and 284--300 keV, thereby generating daily quick-look data products for preliminary analysis.

    \item \rev{The potential flare signals are extracted from the daily quick-look dynamic spectrogram after background subtraction. The intensity of each spectrogram pixel corresponds to the flux at a specific time and energy. The targeted pixels are selected using a flux threshold of $>$0.8 counts $\mathrm{s^{-1}\,keV^{-1}}$. In the resultant spectrogram (see Figure~\ref{fig:img1}(b)), we identify contiguous clusters of four or more spectrogram pixels, such as four consecutive time bins within a single energy bin, or two adjacent energy bins spanning two or more consecutive time bins. Any such cluster (e.g. the red rectangular box in Figure~\ref{fig:img1}(c)) is flagged as a significant detection candidate for subsequent temporal and spectral characterization.}
    
    \item Critical parameters including start/end time, energy range, flux and peak time in each energy band can be derived from these clusters (refer to the inset in Figure~\ref{fig:img1}(c)). Applying thresholds of duration $>$8 s and \rev{total photon counts $>$50 counts $\mathrm{keV^{-1}}$,} clusters satisfying both criteria are classified as genuine flare events.
\end{enumerate}

From 2022 December to 2024 October, HXI detected 2379 solar flares. Classification statistics relative to GOES observations \footnote{\url{https://hesperia.gsfc.nasa.gov/goes/goes_event_listings/}} reveal near-complete detection of X-class flares (61/61, 100\%) and high efficacy for M-class flares (987/1073, 92\%), but significantly lower sensitivity to C-class flares (1331/5034, 26\%). This discrepancy arises primarily from two factors: (1) C-class flares often lack sufficient $>$10 keV emission, and (2) the period of enhanced X-ray emission from a C-class flare coincides with the satellite's passage through the radiation belts or the SAA. \rev{Notably, HXI detected 707 flares above C6-class, accounting for 70\% (707/1031) of GOES-recorded flares in this category.}


\subsection{Flare Selection}
Based on the flare identification results of HXI and GOES, we apply the following criteria to select sample events:
\begin{enumerate}
\item GOES C-class flare, with the corresponding peak 1-8\_$\mathrm{\AA}$\_SXR flux within the range between $\mathrm{10 ^ {-6}}$ and $\mathrm{10 ^ {-5}\ W m^{-2}}$.

\item The background-subtracted peak count rate of the HXI detector D94 in 30--50 keV should be greater than 13.4 $\mathrm{counts\,s^{-1}}$, i.e. larger than the 3$\sigma$ error of the average background count rate $\sim$20 $\mathrm{counts\,s^{-1}}$ for this energy range. The total counts in the flare duration shall be larger than 215 ($>13.4~ \mathrm{counts\,s^{-1}}\times 16$~s) to ensure that the HXR image can be reconstructed with enough counts. 

 
\item The data of the HXI flare samples should not be affected by the particles in the SAA region or Earth's radiation belt.

\item The SXR flux enhancement corresponding to the flares can be clearly identified in the GOES data. 
 
\end{enumerate}

Based on the above criteria, we have selected a total of 127 flares observed from 2022 December 1 to 2024 October 31. Figures~\ref{fig:img1}(d)--(i) show an example of the selected flare. It is important to note that 1118 C-class flares do not meet criterion (2), and we will compare these flares with the selected ones. Additionally, some flares were excluded based on the criterion (3). 

\subsection{Flare Dataset} 

\textbf{\emph{Peak SXR flux of flares}}. \rev{For GOES SXR data, the background level is defined as the local minimum SXR flux within the 30 minutes before the peak of the flare (e.g. red dashed line in Figure~\ref{fig:img1}(d)). }

\label{sec:spectral fitting}
\textbf{\emph{Properties of (non-)thermal electrons in flares}}. For all the selected flares, we used combined spectra data generated from the total flux detectors D92, D93, D94 to obtain X-ray spectra and analyzed the spectra within a time interval of 10--20 s (varying with the count of each flare) around the HXR peak time using the forward fitting method in \texttt{OSPEX}.\footnote{\url{https://hesperia.gsfc.nasa.gov/ssw/packages/spex}}
Each spectrum is fitted with the model including a thermal component from the isothermal bremsstrahlung radiation function (\texttt{f\_vth}) and a nonthermal component from the thick-target bremsstrahlung radiation function with a single power-law electron distribution (\texttt{f\_thick2\_vnorm}), and an albedo correction function is added (an example of the spectral fitting result can be found in Figures~\ref{fig:img1}(f) and (i)). We also examined other combinations of fitting functions, such as multiple thermal components. However, they generally failed to yield robust fits, or the resulting parameters in the fitting do not meet the physical requirements (such as ultra-high-temperature electrons). Although \citet{2008ApJ...673.1181K} applied a multithermal fitting approach to the coronal source spectrum and derived a thermal component with a temperature of approximately 88 MK, they noted that this result pertains to the rapidly evolving coronal source structure.

\textbf{\emph{Location and morphology of flares}}. We utilize the (extreme) ultraviolet ((E)UV) images provided by the Atmospheric Imaging Assembly \citep[AIA;][]{lemenAtmosphericImagingAssembly2012} on board the Solar Dynamic Observatory \citep[SDO;][]{pesnellSolarDynamicsObservatory2012}, combined with HXR imaging of the HXI, to determine the location and morphology of flares, as more than one flare may occur at the same time, resulting in overlapped light curves and multiple brightness enhancements in (E)UV images. The AIA images we used are taken around the impulsive peak time. 
The location of the flare on the solar disk is determined by the centroid of the brightening area in the AIA 1600 $\mathrm{\AA}$ image (such as the black plus sign in Figure~\ref{fig:img1}(h)), which includes pixels with intensity greater than 10\% of the maximum intensity. 

Based on the spectrum fitting results, the fluxes above 20 keV in nearly all flares are dominated by nonthermal radiation. Therefore, we reconstruct HXR images in 20--40 keV \rev{for the same time intervals using the \texttt{HXI\_CLEAN} algorithm and a combination of HXI detectors G3 to G10.}


\section{Results} \label{sec:result}
\subsection{Overall Distribution of Selected Flares}

\begin{figure}[!htbp]
	\centering
	\includegraphics[width=1\textwidth]{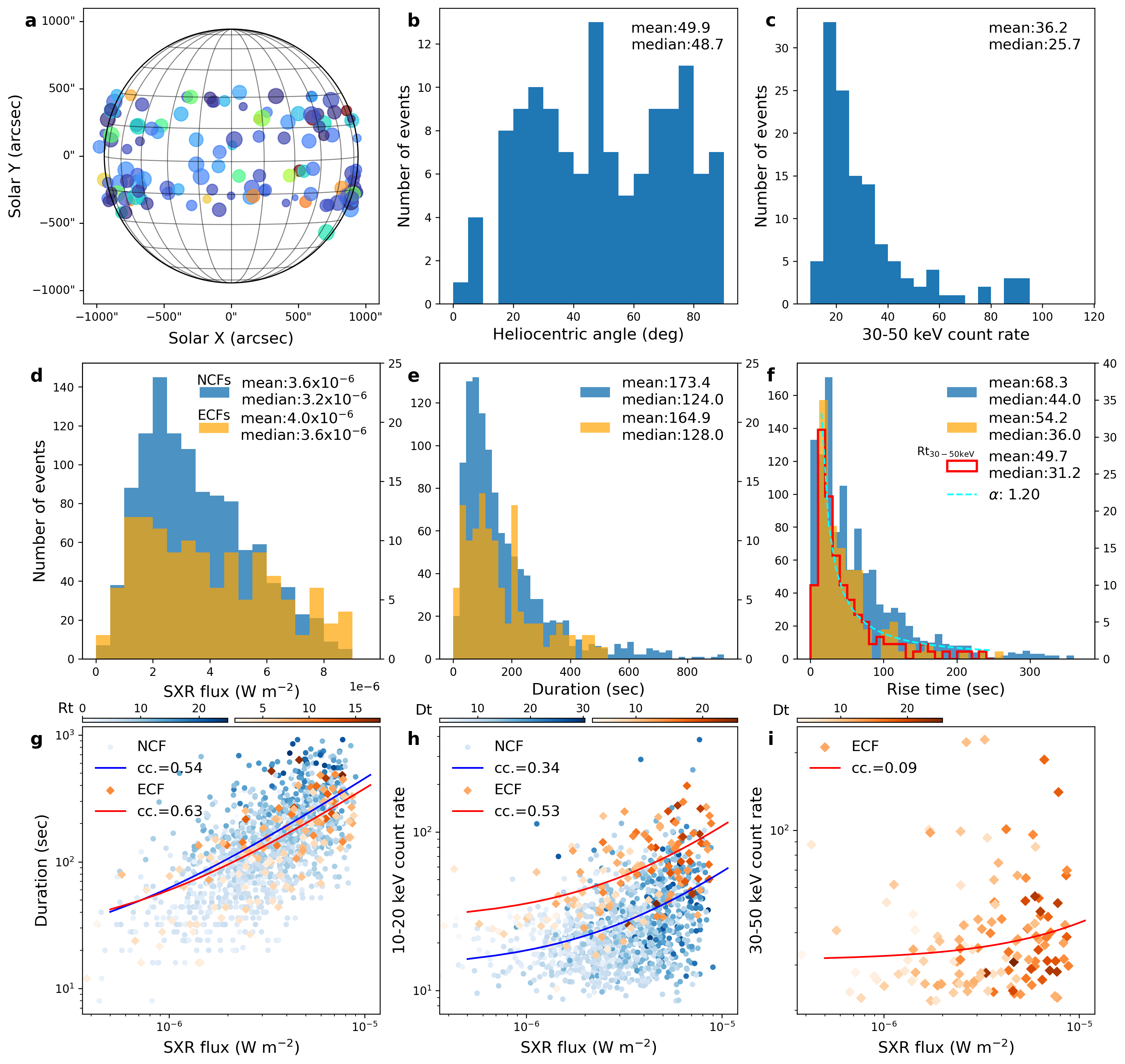} 
	\caption{Distribution and correlation of flare parameters. (a) \rev{The spatial distribution of energetic C-class flares (ECFs)}, where larger scatter points represent larger SXR fluxes, and the color of the scatter points closer to red represents larger HXR count rate. (b) The heliocentric angle distribution of ECFs, which is obtained by transforming the coordinates in panel (a). (c) Statistical distribution of 30--50 keV count rate with ECFs. (d) Distribution of SXR 1--8 $\mathrm{\AA}$ flux, \rev{where dark yellow and blue represent the results of ECFs and normal C-class flares (NCFs)}, respectively. (e) Distribution of duration derived from 10 to 20 keV. (f) Distribution of rise time derived from 10 to 20 keV and 30 to 50 keV (only in ECFs, shown as red step plots). The cyan dashed line represents the power-law fit for the 30--50 keV data, with the corresponding index indicated. (g) Scaling of duration with flare importance. Color-coded scatter points show NCFs (blue) and ECFs (red), with luminance gradients correlating to rise time (a darker hue means a longer rise time). The correlation between the duration and SXR flux with two types of flares is shown in the legend, and the corresponding regression lines are also overlaid on the scatter plot. (h) Scaling of 10--20 keV flux with flare importance. The luminance gradients of points correlate to durations, with darker hues representing longer durations.
    (i) \rev{Scaling of 30--50 keV flux with flare importance. The luminance gradients of points correlate to durations, with darker hues representing longer durations.} } \label{fig:img2}
\end{figure}

Figure~\ref{fig:img2} presents the histogram of the general parameters of the flare samples, and the key information of all the flares is also listed in \rev{Table~\ref{tab:table1}}.
The spatial distribution of flares in Figure~\ref{fig:img2}(a) and the heliocentric angle distribution in Figure~\ref{fig:img2}(b) indicate that ECFs are relatively evenly distributed from the disk center to the limb.
Figure~\ref{fig:img2}(c) shows the flux distribution of 30--50 keV HXR, with mean and median values of 36.2 and 25.7 count $\mathrm{s^{-1}}$, respectively.

Figures~\ref{fig:img2}(d)--(f) present a comparison between ECFs and NCFs in terms of SXR flux, duration, and rise time, which are derived from the 10 to 20 keV energy channel. 
The distribution profiles of SXR flux differ slightly between the two flare types, with ECFs having relatively higher mean and median values compared to NCFs, at $4.0\times 10^{-6}$ and $3.6\times 10^{-6} $ $\mathrm{W m^{-2}}$, respectively.
Regarding the durations, ECFs are mainly concentrated in the range of less than 180 s, with an additional peak around 200 s. Their average and median values are very similar to the NCFs, at 165 and 128 s, respectively.
Since the peak times differ in different energy bands, we also examine the rise time for ECFs in the 30--50 keV channel in Figure~\ref{fig:img2}(f) (shown in the red steps). The profiles of all three variables follow power-law distributions; for example, the fitted spectral index for the rise time distribution in the 30--50 keV channel is 1.2. The rise time for ECFs is shorter than for NCFs, with the high-energy segment having a faster rise time than the low-energy segment. The average and median rise times for ECFs in 30--50 keV are 49.7 and 31.2 s, respectively.

Figures~\ref{fig:img2}(g) and (h) illustrate the duration, rise time, and 10--20 keV emission of ECFs and NCFs as the flare importance changes.
Both types of flares show a positive correlation between duration and SXR flux. The duration of ECFs increases at a slower rate with flare intensity and has a narrower distribution range compared to NCFs. ECFs exhibit significantly higher 10--20 keV flux than NCFs at the same flare importance, while their duration is shorter at the same 10--20 keV flux level. Interestingly, the distribution shown in Figure~\ref{fig:img2}(i) indicates that the 30--50 keV flux of ECFs has no correlation with flare importance (SXR flux) \rev{nor} flare duration. \rev{This differs from the characteristics observed in 10--20 keV emissions and furthermore diverges from the well-established positive correlation between HXR and SXR fluxes reported in previous statistical studies \citep{2008SoPh..250...53S, battagliaSizeDependenceSolar2005}}. It should be emphasized that those earlier correlation analyses incorporated flare events spanning B- to X-class magnitudes, unlike our current statistical sample, which exclusively focuses on C-class flares.

\subsection{Thermal and Nonthermal Properties}

\begin{figure}[!htbp]
	\centering
	\includegraphics[width=1\textwidth]{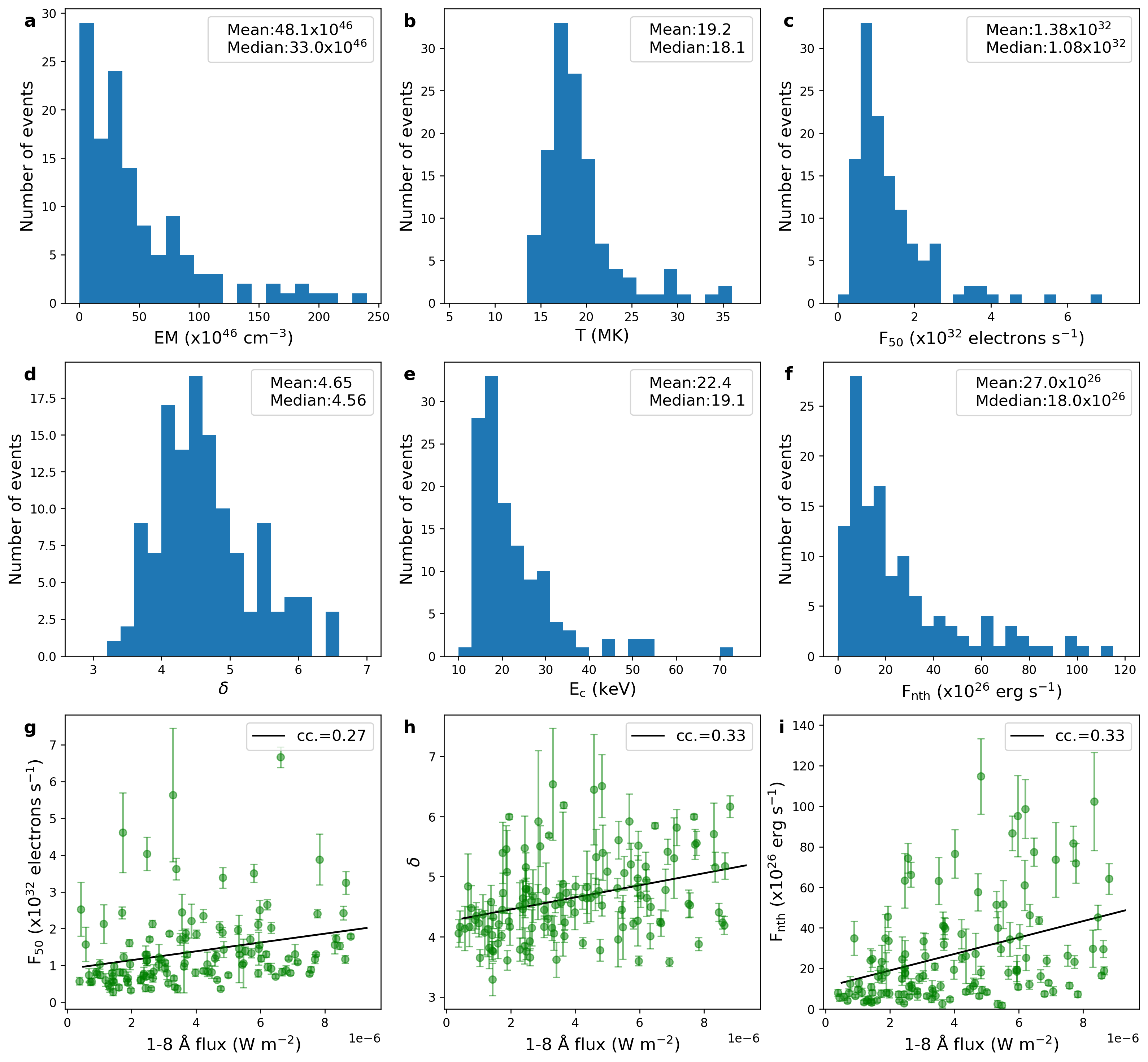} 
	\caption{The distribution and correlation of spectral fitting parameters. (a)--(f) The statistical  distributions of emission measure (EM), temperature ($T$), flux of 50 keV nonthermal electrons ($F_{50}$), nonthermal electron \rev{power-law} index ($\delta$), low-energy cutoff ($E_c$), and nonthermal electron energy flux ($F_{nth}$), with the corresponding mean and median values given in the legends. (g)--(i) The scaling of $F_{50}$, $\delta$ and $F_{nth}$ with SXR flux. Their correlations are given in the legend, and the corresponding regression lines are also overlaid on the scatter plot.} \label{fig:img3}
\end{figure}

\rev{Based on the HXR spectral fitting combining an isothermal model and a thick-target model, as described in Section~\ref{sec:spectral fitting}, we derived the parameters of thermal and nonthermal electrons in all the flare samples and present the statistical results of them in Figure~\ref{fig:img3}}, including emission measures (EMs), temperature ($T$), flux of nonthermal electrons at 50 keV ($F_{50}$), \rev{power-law index of non-thermal electron distribution} ($\delta$), low-energy cutoff ($E_c$), and peak nonthermal energy power ($F_{nth}=\frac{A}{\delta-2}E_{c}^{2-\delta},A =F_{50}/50^{-\delta} $). The electron spectral indices corresponding to each event are listed in \rev{Table~\ref{tab:table1}}.
From Figures~\ref{fig:img3}(a) and (b), we find that the range of temperature distribution of thermal electrons is narrow, concentrated at 20 $\pm$ 6 MK. The EMs of most flares are less than $\mathrm{10 ^ {48}\ cm ^ {-3}}$. The mean/median values of temperature and EM are 19.6/18.1 MK and $\mathrm{4.8 \times 10 ^ {47}/3.3 \times 10 ^ {47}\ cm ^ {-3}}$, respectively. Figures~\ref{fig:img3}(c)--(f) show the properties of nonthermal electrons, with the average and median value of $F_{50}$, $\delta$, $E_c$, and $F_{nth}$ being $\mathrm{1.32 \times 10^{32}/1.08 \times 10 ^ {32}\ electrons\ s^{-1}}$, 4.65/4.56, 22.4/19.1 keV, $\mathrm{2.7 \times 10 ^ {27}/1.8 \times 10 ^ {27} \ erg\ s^{-1}}$, respectively. It is important to note that, because thermal components dominate the low-energy range, the low-energy cutoff derived from \rev{spectral fitting often} represents an upper limit \rev{of $E_c$}.

Similar to the HXR flux distribution, the distributions of EM, $F_{50}$, and $F_{nth}$ also follow an approximate power-law trend. The distribution of $\delta$ is particularly noteworthy. About 33 cases exhibit $\delta$ values greater than 5, resembling typical small flares where electron acceleration efficiency is relatively low. In contrast, the majority of events (94 cases) have $\delta$ values below 5, with a mode of 4.5, aligning with statistical findings for large flares \citep[e.g.,][]{1995ApJ...455..733B,2008SoPh..250...53S}. Moreover, approximately 19 cases have $\delta$ values below 4, a characteristic generally associated with eruptive and impulsive flares \citep[e.g.,][]{warmuthConstraintsEnergyRelease2016a}.
  

The correlation between nonthermal electron parameters and SXR flux is also analyzed. As shown in Figure~\ref{fig:img3}(g)--(i), $F_{50}$ remains relatively constant with increasing SXR flux, while \rev{$\delta$ and $F_{nth}$} exhibit a weak positive correlation with SXR flux \rev{(both correlation coefficients (cc) are 0.33)}. This suggests that in ECFs, lower SXR flux does not correspond to a softer energy spectrum, meaning the electron acceleration efficiency of flares does not decline with decreasing flare importance. 

In the statistical study by \citet{warmuthConstraintsEnergyRelease2016a}, $\delta$ was found to be inversely correlated with SXR flux, indicating that higher SXR flux corresponds to a harder spectrum. This trend has also been observed in microflares \citep{battagliaSizeDependenceSolar2005,2007SoPh..246..339S,hannahRHESSIMicroflareStatistics2008}. However, ECFs exhibit the opposite behavior, with some cases showing spectra as hard as those of X-class flares. This characteristic closely resembles that of recently identified hard microflares \citep{2024A&A...683A..41S,2024A&A...691A.172B,2024ApJ...964..142A,lysenkoStatisticsColdEarly2018}. Furthermore, the lack of correlation between nonthermal electron flux and SXR flux, as well as the previously mentioned absence of a correlation between HXR and SXR flux, further suggests that the energy transport process of accelerated electrons in these flares differs fundamentally from that of conventional flares.
 
\subsection{Relationship between Flares and Other Features}

In order to better understand the characteristics of ECFs, we incorporate the observational features of other wave bands to analyze the energy release and transport processes. Specifically, we examine the correlation between flares and CMEs, jets, and low-frequency type III bursts. The data involved come from SDO/AIA, SOHO/LASCO \citep{1995SoPh..162....1D,1995SoPh..162..357B}, and WIND/WAVES \citep{1995SSRv...71..231B}.

\begin{figure}[!htbp]
	\centering
	\includegraphics[width=1\textwidth]{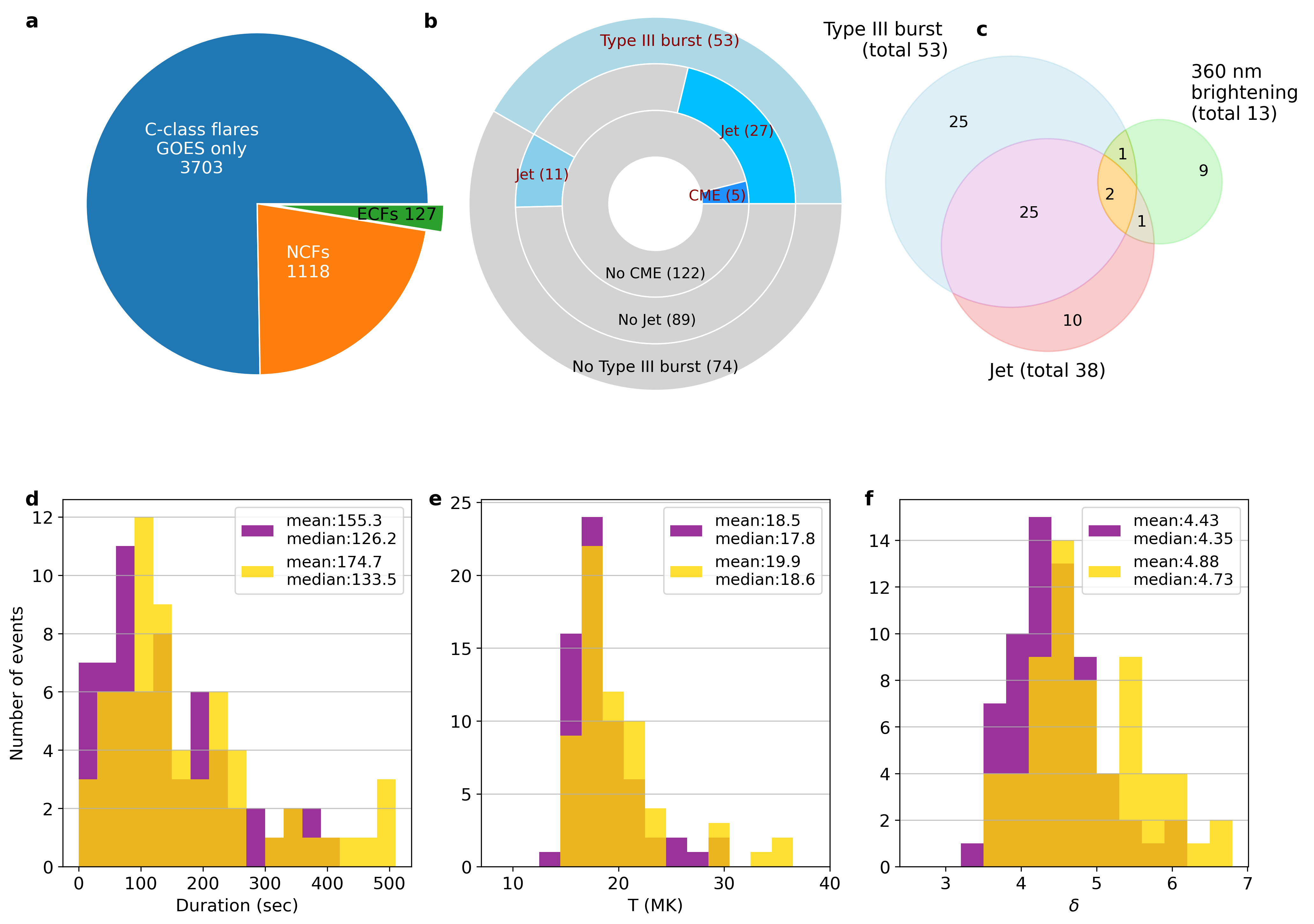} 
	\caption{\rev{The correlation between energetic C-class flares (ECFs) and other phenomena.} (a) \rev{The proportion of ECFs and normal C-class flares (NCFs) in C-class flares.} (b) Correlation between ECFs and CMEs, jets, and type III bursts. The presence or absence of relevant phenomena is indicated in blue and gray, respectively, and their quantities are marked accordingly. The pie chart represents the intersection of various phenomena from the outside to the inside, with different color depths. For example, a dark blue color in the center of the chart indicates that five CMEs occur simultaneously with both jets and type III bursts. (c) Correlation between jets, type III bursts, and \rev{360 nm brightening} in ECFs. (d)--(f) The statistical distribution between the duration, temperature, and \rev{nonthermal electron power-law} index corresponding to ECFs with and without an open magnetic field. The purple and gold bars indicate the presence and absence of an open magnetic field, respectively.} \label{fig:img4}
\end{figure}

Out of a total of 127 events, 53 were associated with type III bursts, 38 with jets, and only five with CMEs. These observational features are not independent of one another. 
As shown in Figure~\ref{fig:img4}(b), 27 flares were accompanied by both type III bursts and jets, with all five CME events included within this subset.
Previous studies on the association between CMEs and flares have indicated that the likelihood of a flare being linked to a CME increases with flare importance \citep{yashiroVisibilityCoronalMass2005,2007ApJ...665.1428W}, with approximately 20\% of C-class and 49\% of M-class flares associated with CMEs. In contrast, our sample shows \rev{the association rate is less than 5\% for ECFs}, significantly lower than previous statistical findings. 

\begin{figure}[!htbp]
	\centering
	\includegraphics[width=1\textwidth]{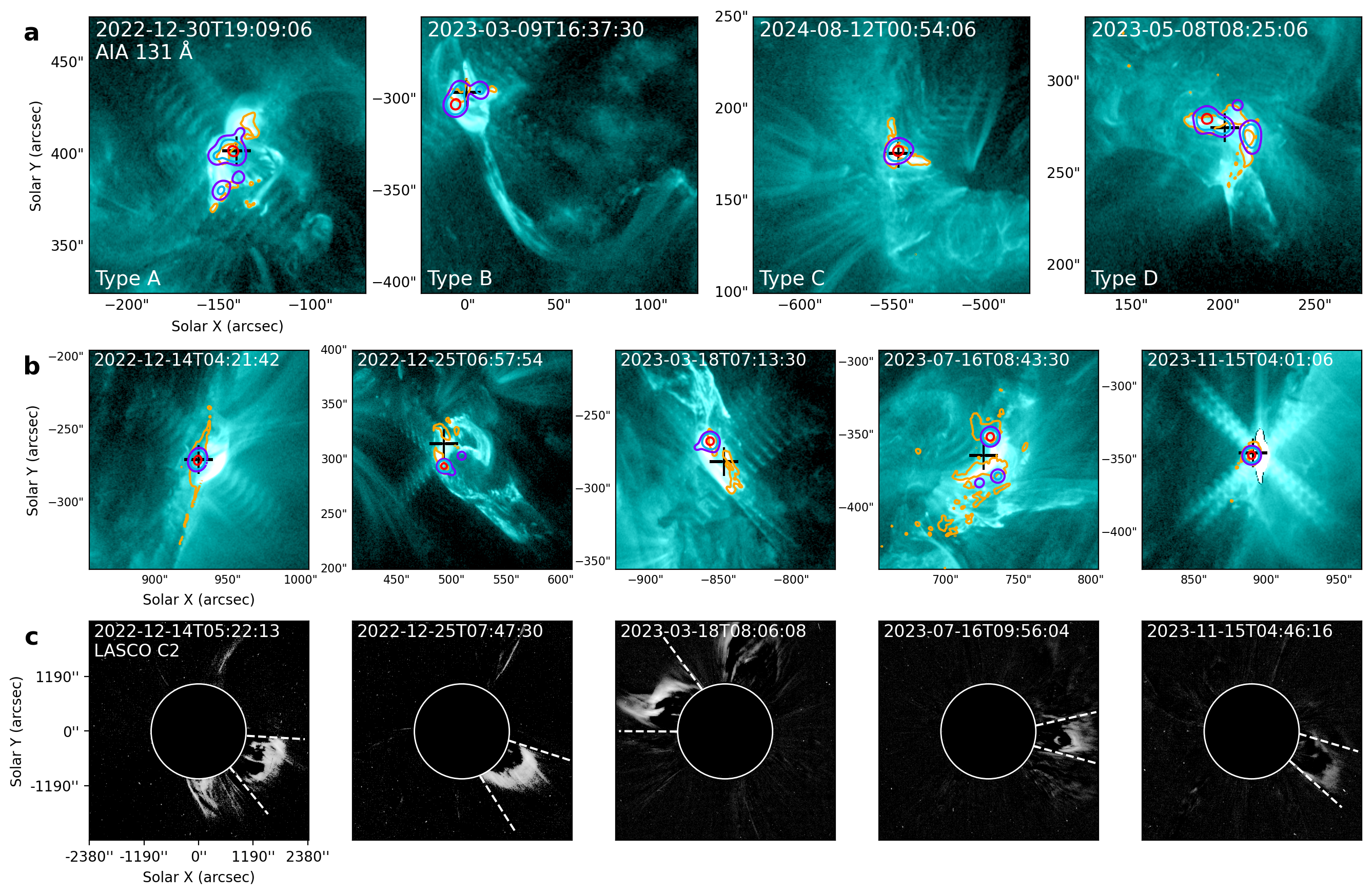} 
	\caption{The morphology of different types of flares. The top panel (a) displays four cases of flares without accompanying CMEs, each representing a distinct category (types A, B, C, and D). The specific characteristics of each category are detailed in the main text. The middle panel (b) presents all five cases of flares associated with CMEs. The background images in panels (a) and (b) are both AIA 131 $\mathrm{\AA}$ images, while the rainbow contour lines in red, cyan, and purple represent 75\%, 30\%, and 15\% of the peak intensity in HXI 20–-40 keV imaging, respectively. The orange contour indicates the 10\% maximum intensity in the AIA 1600 $\mathrm{\AA}$ image, with the centroid corresponding to the flare ribbons marked with a black plus sign. The bottom panel (c) corresponds to panel (b) and shows the CME observation from LASCO C2. The two dashed lines roughly indicate the ejection direction of the CME.} \label{fig:img5}
\end{figure}

EUV and WL observations (Figures~\ref{fig:img5}(b) and (c)) reveal that the morphology of these five CME events is characterized by narrow CMEs driven by jets propagating outward along open magnetic field lines, rather than the typical three-part CMEs associated with filament or MFR eruptions \citep[see the review of][]{chenCoronalMassEjections2011}. Specifically, prior to the five flares, a narrow and slow-moving jet was observed along the open magnetic field lines. As the flare reaches its impulsive phase, a broader and faster jet is rapidly ejected in the same direction, ultimately forming a CME. 

Additionally, Figure~\ref{fig:img5}(a) illustrates four types of flares that are not associated with CMEs: fully confined flares without jets or type III bursts (type A, 63 cases), flares accompanied solely by jets (type B, 11 cases), flares associated only with type III bursts (type C, 26 cases), and flares associated with both jets and type III bursts but without CMEs (type D, 27 cases).

\rev{Furthermore, numerous studies have demonstrated that WL emission is both temporally and spatially correlated with HXR radiation \citep[e.g.,][]{hudsonWhiteLightFlaresTRACE2006a,kruckerCOSPATIALWHITELIGHT2015}, highlighting their strong association with energetic electrons. Recently, a statistical comparison of WL flares at 360 nm and 617.3 nm reveals a close spatiotemporal correspondence between the two wavelengths \citep{2024ApJ...972L...1L}. Therefore, observations at 360 nm can meaningfully track the behavior of WL flares. Given the high electron acceleration efficiency in ECFs, we also utilize the data (full-disk image at 360 nm with a 2 minute cadence) from the White-light Solar Telescope (WST) on the Ly$\alpha$ Solar Telescope \citep{2019RAA....19..158L,2024SoPh..299..118C} on board ASO-S to determine whether the selected flares exhibit 360 nm brightenings. According to the method described in \citet{jing2024SoPh..299...11J}, significantly brightening areas are identified by applying a threshold above the quiet-region intensity, which is defined as the average intensity within the 10 minutes before flare onset and after its end. In this study, the threshold is set at 8\%, which exceeds 3 times the background fluctuation (less than 2\%)}


Among the analyzed flares (118 events, excluding nine due to data gaps in WST), concurrent \rev{360 nm brightenings were observed in 13 cases. The correlation between these flares and the previously mentioned features is illustrated in Figure~\ref{fig:img4}(c).
Of these 13 events, none were related to CMEs, two were simultaneously linked to both type III bursts and jets, one was associated only with a type III burst, and another solely with a jet. The remaining nine cases were not related to any other features.
Statistical studies have shown that the incidence of WL flares increases with flare intensity \citep{2018ApJ...867..159S,2020ApJ...904...96C,jing2024SoPh..299...11J}. The probability of WL enhancement in C-class flares is around 10\%, which is consistent with the 11\% incidence observed in ECFs at 360 nm. Notably, except for four flares with $\delta > 5$ (5.21--6.45), the remaining indices range from 3.6 to 4.83. This suggests that 360 nm brightening is more likely to occur in events involving high-energy accelerated electrons. }

Previous statistical studies on the correlation between type III bursts and HXR emissions have shown that a harder X-ray spectrum corresponds to a stronger correlation \citep{hamiltonStatisticalStudyCorrelation1990}. To verify whether this holds in our sample, we compared the properties of electrons in \rev{the flares that are likely associated} with closed magnetic fields (i.e., type A flares) to those in flares \rev{that are likely associated} with open magnetic fields (types B, C, and D). The distributions of selected parameters are shown in Figures~\ref{fig:img4}(d)--(f).  
The results indicate that the average and median durations of \rev{the latter group} are slightly shorter than those of \rev{the former group}, and the temperature \rev{of the latter group} is marginally lower.
Additionally, the \rev{distribution of $\delta$ of the latter group} exhibits an overall shift toward lower values, with an average of 4.43 compared to 4.88 for \rev{the former group with closed field}. This finding is consistent with the conclusion that flares with harder spectra exhibit a stronger correlation with type III bursts.
Other parameters, such as HXR/SXR flux, emission measure, and nonthermal energy flux, show no significant differences between the two categories and are therefore not included in Figure~\ref{fig:img4}.

\section{Discussion and Summary} \label{sec:summary}

In this study, we investigate a set of C-class solar flares with strong HXR emissions above 30 keV, identified through HXI and GOES observations from 2022 December to 2024 October. The statistical analysis highlights several key characteristics of these ECFs, including:
\begin{enumerate}
    \item For the same peak flux in the 10--20 keV range, ECFs have significantly shorter durations than NCFs, and their corresponding SXR peak flux is also notably lower.
    \item X-ray spectral analysis reveals that ECFs have a harder spectrum compared to NCFs, which typically feature very steep spectra or are dominated by thermal components. The average \rev{power-law} index of nonthermal electrons in ECFs is 4.65, suggesting a high electron acceleration efficiency in these flares.    
    \item In ECFs, there is no significant correlation between HXR flux, nonthermal electron flux, nonthermal electron \rev{power-law} index, and SXR flux. This contrasts with typical flares, where thermal and nonthermal emissions \citep{battagliaSizeDependenceSolar2005,2008SoPh..250...53S,hannahRHESSIMicroflareStatistics2008}, as well as thermal and nonthermal electrons \citep{warmuthConstraintsEnergyRelease2016a,warmuthConstraintsEnergyRelease2016}, are usually correlated.
    \item Among the 127 ECFs, only five are associated with CMEs, which is significantly lower than the typical association ratio of approximately 20\% between flares and CMEs for C-class flares in previous statistical studies. Besides, all five CMEs are narrow CMEs produced by jets.
    \item Half of the ECFs (64/127) exhibit open magnetic structures, which are associated with harder spectra. Additionally, 11\% of the ECFs show \rev{360 nm brightenings}, and most of these flares have spectra harder than the average ECFs.
\end{enumerate}
These characteristics distinguish them from normal C-class flares and may provide insights into the underlying energy release and transport processes.

The most distinctive aspects of ECFs are their weak thermal response and highly efficient electron acceleration process.
The thermal radiation in flares primarily originates from high-temperature plasma within the flare loops, which are filled by evaporated plasma driven by the heating of accelerated electron beams during the impulsive phase. Based on the upward velocity of the evaporated plasma flow, the evaporation process can be classified into two types: explosive \citep{2006ApJ...638L.117M} and gentle \citep{2006ApJ...642L.169M} evaporations. The energy density of the nonthermal electron beam affects the evaporation speed. Generally, when the energy injection rate of nonthermal electrons is below $3 \times 10^{10} \,\mathrm{erg\,cm^{-2}\,s^{-1}}$, a gentle evaporation process is expected (see the review by \citet{fleishmanCOLDTENUOUSSOLAR2011} and references therein).  

In ECFs, the average nonthermal energy flux is 
$2.7 \times 10^{27} \,\mathrm{erg\,s^{-1}}$. According to the HXR imaging results in 20--40 keV, we assume an injection area of $8^{\prime\prime}\times8^{\prime\prime}$ and estimated the average nonthermal energy injection rate to be $7.87\times 10^{9} \,\mathrm{erg\,cm^{-2}\,s^{-1}}$, which is insufficient to drive explosive evaporation. Moreover, simulations by \citet{2015ApJ...808..177R} suggest that compared to high-energy electrons, low-energy electrons are more effective in heating the chromosphere, particularly in the case of gentle evaporation. Therefore, the harder spectra of ECFs are less favorable for chromospheric heating. 
Additionally, the high-energy radiation in most ECFs lasts less than 1 minute. These facts suggest that the accelerated electrons in ECFs may not be sufficient to \rev{produce} a significant amount of high-temperature plasma and produce strong SXR emissions.

\rev{HXR imaging reveals that most footpoint sources appear as either a single compact source or closely distributed double sources, indicating that the impact area of nonthermal electrons on the chromosphere is highly localized. Additionally, the flare ribbons observed in the 1600 $\mathrm{\AA}$ band are relatively small and close. These characteristics suggest that these flare loop systems have limited volume.} 


\rev{A distinguishing feature of ECFs is their weak association with CMEs, suggesting that they may not confirm the standard flare model. In that framework, the eruption of an MFR drives a CME and powers a flare through magnetic reconnection underneath the MFR, which heats plasma and accelerates electrons. By contrast, the main acceleration mechanism of electrons in ECFs may differ from the standard model. Another possible reason is that the magnetic field gradients/strength in confined regions may be stronger than those in eruptive cases.}


Recent statistical studies on hard microflares have shown that these events, which exhibit efficient electron acceleration, consistently have at least one footpoint located within a sunspot, suggesting that strong magnetic fields play a crucial role in efficient electron acceleration \citep{2024A&A...691A.172B}. Following this, we examined the footpoint locations of ECFs and found that over three-quarters of the flares (94 out of 121, excluding six out-of-disk events) had at least one footpoint anchored in a sunspot, consistent with the characteristics of hard microflares. This suggests that strong magnetic fields are linked to efficient electron acceleration across flares of different magnitudes. They may influence the electric field in the magnetic reconnection process, thereby enhancing electron acceleration. However, to fully understand this connection, it is essential to investigate the actual magnetic field configuration of flares and how electron acceleration evolves within it, making this a key focus for future studies.

Since these flares exhibit strong HXR emission but weak SXR emission, the GOES-defined classification may not adequately reflect the actual properties of the accelerated electrons. This highlights the need for a more refined understanding of solar flares and the development of additional classification schemes, such as one based on HXR flux, to better represent flare intensity.

\begin{acknowledgments}
The work is supported by the National Key R\&D Program of China 2022YFF0503002, and the NSFC 12333010, 11921003, 11820101002. The ASO-S mission is supported by the Strategic Priority Research Program on Space Science, the Chinese Academy of Sciences, grant No. XDA15320000. SDO is a mission for NASA's Living with a Star program. 
\end{acknowledgments}

\bibliography{main_v2.5}{}
\bibliographystyle{aasjournal}

\begin{table}[ht]
	\centering
        \tiny

	\begin{threeparttable}
            \caption{Information of 127 ECFs}
            \label{tab:table1}%
		\begin{tabular}{llccccc|llccccc} 
		\toprule
		Peak Time &\parbox{0.3cm}{GOES\\Class} & \parbox{0.3cm}{$\delta$} & \parbox{0.75cm}{360 nm\\-B\tnote{*}} & \parbox{0.3cm}{CME}   & \parbox{0.3cm}{Jet}  & \parbox{0.9cm}{Type III\\Burst}  & Peak Time & \parbox{0.3cm}{GOES\\Class} & \parbox{0.3cm}{$\delta$} & \parbox{0.75cm}{360 nm\\-B\tnote{*}} & \parbox{0.3cm}{CME}   & \parbox{0.3cm}{Jet} & \parbox{0.9cm}{Type III\\Burst} \\
		\midrule
		2022-12-02T09:18:48 & C8.5  & 5.54  & Y     & N     & N     & N     & 2023-11-23T07:12:16 & C6.9  & 3.77  & --     & N     & N     & N \\
		2022-12-14T04:09:36 & C5.9  & 4.71  & --    & Y     & Y     & Y     & 2023-11-30T15:42:09 & C2.2  & 4.07  & --    & N     & N     & Y \\
		2022-12-16T19:06:07 & C9.1  & 4.94  & --     & N     & N     & Y     & 2023-12-01T11:56:21 & C5.2  & 3.9   &  --    & N     & N     & Y \\
		2022-12-18T09:44:01 & C2.5  & 3.29  & --     & N     & Y     & Y     & 2023-12-21T21:20:31 & C4.2  & 4.58  & --     & N     & Y     & N \\
		2022-12-20T04:09:28 & C4.6  & 4.53  & --     & N     & N     & Y     & 2023-12-29T07:52:24 & C9.8  & 6.17  & --    & N     & N     & Y \\
		2022-12-20T13:53:44 & C6.6  & 5.61  & --      & N     & N     & N     & 2023-12-30T15:18:37 & C6.1  & 4.76  & --    & N     & N     & N \\
		2022-12-24T04:10:53 & C4.1  & 4.17  & --     & N     & Y     & Y     & 2024-01-01T22:40:02 & C1.6  & 4.14  & --     & N     & N     & Y \\
		2022-12-25T06:59:13 & C7.8  & 3.58  & --      & Y     & Y     & Y     & 2024-01-03T10:15:05 & C3.0  & 4.64  & --     & N     & N     & N \\
		2022-12-27T20:35:41 & C4.1  & 4.73  & --    & N     & N     & N     & 2024-01-11T01:09:24 & C5.2  & 6     & --     & N     & N     & N \\
		2022-12-30T19:08:35 & C4.4  & 5.92  & --     & N     & N     & N     & 2024-02-04T15:15:33 & C4.5  & 4.51  & --     & N     & Y     & N \\
		2023-01-10T14:46:32 & C9.1  & 4.65  & --     & N     & Y     & N     & 2024-02-04T21:06:43 & C4.6  & 5.4   & --     & N     & N     & N \\
		2023-01-24T18:50:00 & C3.8  & 3.61  & --    & N     & N     & N     & 2024-02-06T06:56:12 & C6.5  & 4.17  & --      & N     & N     & N \\
		2023-02-05T03:24:38 & C6.7  & 3.6   & Y     & N     & Y     & Y     & 2024-02-06T12:51:27 & C9.4  & 4.27  & --     & N     & N     & N \\
		2023-02-10T10:50:20 & C9.9  & 4.78  & --     & N     & N     & N     & 2024-02-09T07:30:01 & C9.7  & 5.71  & --     & N     & N     & N \\
		2023-02-18T18:34:43 & C7.9  & 4.69  & --     & N     & Y     & Y     & 2024-02-11T00:25:54 & C6.0  & 4.13  & --     & N     & N     & N \\
		2023-02-23T04:43:11 & C4.5  & 3.82  & --      & N     & Y     & Y     & 2024-02-14T17:55:13 & C7.7  & 4.22  & --     & N     & N     & N \\
		2023-02-23T08:09:14 & C7.2  & 3.96  & --      & N     & N     & Y     & 2024-02-19T12:45:22 & C6.0  & 6.45  & Y     & N     & N     & N \\
		2023-03-02T04:44:34 & C9.3  & 6     & --      & N     & N     & N     & 2024-02-23T23:12:18 & C5.0  & 5.08  & --    & N     & Y     & Y \\
		2023-03-04T02:24:58 & C6.0  & 4.88  & --     & N     & Y     & Y     & 2024-02-24T08:09:13 & C4.5  & 4.45  & --     & N     & Y     & N \\
		2023-03-09T16:37:31 & C2.2  & 4.17  & --     & N     & Y     & N     & 2024-02-24T09:09:33 & C5.3  & 6.19  & --     & N     & N     & N \\
		2023-03-12T06:07:16 & C1.3  & 4.06  & --     & N     & N     & N     & 2024-02-25T09:06:58 & C2.5  & 4.36  & --     & N     & N     & N \\
		2023-03-18T07:13:03 & C9.5  & 4.25  & --     & Y     & Y     & Y     & 2024-03-01T01:24:07 & C4.0  & 4.6   & --     & N     & N     & N \\
		2023-03-19T02:14:52 & C3.1  & 4.02  &  N/A     & N     & N     & N     & 2024-03-07T12:02:23 & C6.4  & 4.45  & --     & N     & N     & N \\
		2023-03-19T18:07:51 & C5.1  & 4.84  & --     & N     & Y     & Y     & 2024-03-15T05:14:18 & C5.6  & 4.23  & --     & N     & Y     & Y \\
		2023-03-21T13:53:35 & C5.5  & 4.65  & --     & N     & N     & N     & 2024-03-25T22:29:27 & C5.1  & 4.15  & --    & N     & Y     & Y \\
		2023-03-31T20:52:19 & C9.8  & 4.19  & --     & N     & Y     & Y     & 2024-03-26T02:25:56 & C5.9  & 4.73  & --     & N     & Y     & Y \\
		2023-04-10T13:49:24 & C6.0  & 6.51  & --     & N     & N     & N     & 2024-03-26T19:50:19 & C6.9  & 4.83  & --     & N     & N     & N \\
		2023-04-13T10:40:14 & C8.5  & 5.42  & --     & N     & N     & Y     & 2024-03-29T00:52:59 & C4.9  & 4.15  & --     & N     & N     & Y \\
		2023-04-16T17:36:32 & C9.0  & 4.53  & Y     & N     & Y     & N     & 2024-04-18T11:49:35 & C9.0  & 5.4   & --     & N     & N     & N \\
		2023-04-18T05:28:50 & C2.4  & 4.58  & --     & N     & Y     & N     & 2024-04-25T01:31:33 & C8.7  & 5.85  & --    & N     & N     & N \\
		2023-05-08T08:23:48 & C3.1  & 4.45  & --      & N     & Y     & Y     & 2024-05-03T18:52:25 & C7.5  & 4.16  & --     & N     & N     & Y \\
		2023-05-08T14:18:27 & C9.7  & 5.18  & --      & N     & N     & Y     & 2024-05-05T02:59:54 & C8.8  & 3.62  & --     & N     & N     & N \\
		2023-05-19T19:10:35 & C5.2  & 4.24  & --    & N     & N     & Y     & 2024-05-06T17:07:08 & C4.4  & 4.3   & --     & N     & N     & N \\
		2023-05-20T04:43:04 & C6.1  & 5.09  & --     & N     & N     & N     & 2024-05-06T18:21:03 & C3.7  & 3.93  &  --    & N     & N     & N \\
		2023-05-22T13:26:24 & C7.5  & 4.83  & Y    & N     & N     & N     & 2024-05-12T00:23:38 & C6.5  & 3.79  & --     & N     & Y     & Y \\
		2023-05-27T04:34:06 & C8.7  & 3.88  & Y     & N     & N     & Y     & 2024-05-13T01:07:58 & C9.5  & 5.06  & --     & N     & N     & N \\
		
		2023-06-09T14:37:38 & C4.5  & 6.54  & --     & N     & N     & N     & 2024-05-18T20:14:53 & C4.0  & 4.56  & --    & N     & N     & N \\
		2023-06-19T03:40:48 & C4.4  & 4.46  & N/A     & N     & N     & N     & 2024-05-31T03:33:33 & C9.8  & 5.56  &  --    & N     & N     & N \\
		2023-06-20T11:01:45 & C8.6  & 5.32  & --      & N     & N     & N     & 2024-06-04T03:47:08 & C8.4  & 5.82  & Y     & N     & N     & N \\
		2023-06-25T21:58:05 & C7.0  & 5.33  & --     & N     & Y     & Y     & 2024-06-21T12:38:32 & C4.2  & 5.46  & --    & N     & N     & N \\
		2023-07-02T08:49:03 & C8.9  & 4.81  & --     & N     & N     & N     & 2024-07-11T15:51:43 & C4.1  & 5.48  & --     & N     & N     & N \\
		2023-07-05T18:15:25 & C4.4  & 4.68  & --     & N     & N     & N     & 2024-07-16T13:03:25 & C5.0  & 4.8   &   N/A    & N     & N     & N \\
		2023-07-11T02:13:41 & C8.8  & 4.51  &   N/A    & N     & N     & N     & 2024-07-22T06:55:23 & C6.2  & 4.54  & --    & N     & N     & N \\
		2023-07-15T16:01:30 & C3.4  & 3.85  & --     & N     & Y     & N     & 2024-08-02T17:11:30 & C6.1  & 4.21  & --     & N     & N     & N \\
		2023-07-16T08:37:09 & C5.0  & 4.06  & --     & Y     & Y     & Y     & 2024-08-03T02:55:33 & C9.3  & 4.68  & --     & N     & N     & N \\
		2023-07-16T17:16:36 & C5.8  & 4.64  & --    & N     & N     & N     & 2024-08-06T16:13:32 & C5.8  & 5.51  & --    & N     & N     & N \\
		2023-07-18T22:42:17 & C8.6  & 5.52  & --    & N     & N     & N     & 2024-08-07T02:17:56 & C5.7  & 3.89  & --     & N     & N     & Y \\
		2023-07-26T15:20:36 & C8.1  & 4.97  & --    & N     & N     & Y     & 2024-08-11T02:54:13 & C6.8  & 4.8   & --     & N     & N     & Y \\
		2023-07-31T11:48:49 & C3.1  & 3.77  & --     & N     & Y     & N     & 2024-08-12T00:52:49 & C6.1  & 3.76  & --     & N     & N     & Y \\
		2023-08-06T01:10:16 & C5.3  & 4.84  & --    & N     & Y     & Y     & 2024-08-13T07:33:40 & C6.7  & 5.16  &  --   & N     & N     & Y \\
		2023-08-26T21:48:32 & C2.6  & 4.26  & --     & N     & Y     & Y     & 2024-08-13T08:03:30 & C7.3  & 5.69  & --     & N     & N     & Y \\
		2023-09-06T00:49:40 & C5.2  & 4.66  & --     & N     & N     & Y     & 2024-08-14T17:21:51 & C9.5  & 4.28  & --     & N     & N     & N \\
		2023-09-06T17:55:20 & C5.6  & 3.79  & --     & N     & N     & N     & 2024-08-22T15:27:31 & C7.8  & 5.21  &  Y   & N     & N     & N \\
		2023-09-16T15:50:39 & C2.7  & 4.33  & --     & N     & Y     & Y     & 2024-08-25T13:09:37 & C2.5  & 4.17  & --     & N     & N     & Y \\
		2023-09-22T06:59:04 & C6.5  & 3.66  & --     & N     & Y     & Y     & 2024-09-06T05:54:34 & C7.8  & 4.52  & Y    & N     & N     & N \\
		2023-09-30T11:55:29 & C9.6  & 5.16  & --     & N     & N     & N     & 2024-09-06T05:59:58 & C7.1  & 4.41  & --     & N     & N     & N \\
		2023-10-01T03:23:24 & C9.6  & 4.25  & Y     & N     & N     & N     & 2024-10-02T18:29:49 & C7.1  & 4.76  & --    & N     & N     & Y \\
		2023-10-09T11:34:52 & C7.5  & 5.17  & --    & N     & N     & Y     & 2024-10-03T06:46:01 & C8.7  & 4.01  & Y     & N     & Y     & Y \\
		2023-10-11T09:37:10 & C7.4  & 4.01  & Y     & N     & N     & N     & 2024-10-04T13:43:34 & C9.4  & 3.65  &  N/A     & N     & N     & Y \\
		2023-11-05T21:33:21 & C9.6  & 4.41  & Y     & N     & N     & N     & 2024-10-05T06:13:30 & C6.6  & 4.03  & --     & N     & Y     & Y \\
		2023-11-11T03:57:19 & C7.6  & 4.23  &  N/A     & N     & Y     & N     & 2024-10-10T00:12:49 & C6.1  & 4.37  & --    & N     & N     & Y \\
		2023-11-15T03:59:28 & C8.5  & 4.55  &  N/A     & Y     & Y     & Y     & 2024-10-14T03:06:19 & C2.1  & 4.48  & --     & N     & Y     & Y \\
		2023-11-19T08:31:38 & C7.8  & 5.92  &  N/A    & N     & N     & Y     & 2024-10-22T12:09:57 & C5.1  & 4.78  & --     & N     & N     & N \\
		2023-11-20T10:13:48 & C5.1  & 4.57  &  N/A    & N     & Y     & N     &     --  &  --     &  --     &   --    &   --    &   --    & -- \\
		\bottomrule 
    	\end{tabular}%
    	\begin{tablenotes}
            \footnotesize
    	\item[*] In the table,``Y'' and ``N'' represent the presence or absence of corresponding phenomena, respectively. However, there is a slight difference in the labeling for the \rev{360 nm-B (360 nm brightening) column, and its classification is based on observations from ASO-S/WST. Due to observation gaps, some flares may be missing,} in which case the cell will be filled with \rev{``N/A''}. Additionally, it is important to note that the time resolution of the WST in normal mode is 2 minutes. However, the duration of radiation above 30 keV in many flares is much shorter than 2 minutes, which means WST may miss some periods of 360 nm brightening. Consequently, the number identified here should be considered as a lower limit. To differentiate these cases from other phenomena, we use ``--'' to indicate events where no 360 nm brightening was detected due to the 2-minute time resolution.
    	\end{tablenotes}
 
    \end{threeparttable}

\end{table}%

\end{document}